\documentclass{ifacconf}

\usepackage{graphics}
\usepackage{amsmath,amssymb,amsfonts}
\usepackage{bm}
\usepackage{algorithm}
\usepackage{algpseudocode}
\usepackage[dvipsnames]{xcolor}
\usepackage{natbib}
\usepackage{siunitx}
\usepackage{booktabs}
\usepackage{tikz}
\usepackage{pgfplots}
\usepackage{etoolbox}
\usepackage{atbegshi}
\usepackage{microtype}

\apptocmd{\thebibliography}{\small\setlength{\itemsep}{0pt}\setlength{\parsep}{0pt}}{}{}

\pgfplotsset{compat=1.18}
\usetikzlibrary{arrows,shapes,backgrounds,patterns,fadings,decorations,
	positioning,external,arrows.meta,pgfplots.fillbetween, pgfplots.groupplots}

\newtheorem{assumption}{Assumption}
\newtheorem{remark}{Remark}

\algrenewcommand\algorithmicrequire{\textbf{Input:}}
\algrenewcommand\algorithmicensure{\textbf{Output:}}

\usepackage[bookmarks=false, implicit=false]{hyperref}

\newcommand\copyrighttext{%
	\footnotesize \textcopyright 2026 the authors. This work has been accepted to IFAC for publication under a Creative Commons Licence CC-BY-NC-ND.}

\AtBeginShipoutFirst{%
	\begin{tikzpicture}[remember picture,overlay]
		\node[anchor=south,yshift=65pt] at (current page.south) {\fbox{\parbox{\dimexpr\textwidth-\fboxsep-\fboxrule\relax}{\copyrighttext}}};
	\end{tikzpicture}%
}

\begin{document}

\begin{frontmatter}

\title{Learning Dynamics from Infrequent Output Measurements for Uncertainty-Aware Optimal Control\thanksref{footnoteinfo}} 

\thanks[footnoteinfo]{Funded by the Deutsche Forschungsgemeinschaft (DFG, German Research Foundation) -- GRK 3081/1 -- Project number 534429653.}

\author{Robert Lefringhausen, Theodor Springer, Sandra Hirche}

\address{Chair of Information-oriented Control, School of Computation, Information and Technology, Technical University of Munich, Germany (e-mail:\{robert.lefringhausen, theo.springer, hirche\}@tum.de)}

\begin{abstract}  
Reliable optimal control is challenging when the dynamics of a nonlinear system are unknown and only infrequent, noisy output measurements are available. This work addresses this setting of limited sensing by formulating a Bayesian prior over the continuous-time dynamics and latent state trajectory in state-space form and updating it through a targeted Metropolis--Hastings sampler equipped with a numerical ODE integrator. The resulting posterior samples are used to formulate a scenario-based optimal control problem that accounts for the uncertainty in the dynamics and latent state and is solved using standard nonlinear programming methods. The approach is validated in a numerical case study on glucose regulation using a Type~1 diabetes model.
\end{abstract}


\end{frontmatter}


\section{Introduction}

Accurate dynamical models are fundamental for the optimal control of nonlinear systems. Although first-principles models may describe the general structure of many systems, important parameters or effects often remain unknown, limiting their direct use for control. This has sparked increasing interest in data-driven approaches that compensate for such limitations by relying on data collected from the dynamical system. Yet, in many real-world applications such as robotics or biological systems, obtaining a reliable model from data is often challenging: available measurements are generally limited, only outputs may be accessible rather than full-state information, sampling may be infrequent due to hardware or communication constraints, and the data is typically corrupted by noise. Under these conditions, models learned from data remain subject to substantial uncertainty, and controllers must explicitly account for it to ensure reliable performance.

To account for uncertainty in learned dynamics, control approaches often rely on Bayesian formulations such as Gaussian process state-space models, which provide a principled way to quantify the effects of limited data and measurement noise \citep{scampicchio2025}. Moreover, by combining available measurements with prior knowledge, these models regularize the inference, thereby improving prediction reliability when observations are scarce and noisy. However, existing Bayesian approaches fall short in settings with infrequent and partial measurements. Bayesian discrete-time state-space models \citep{umlauft2018} require frequent full-state measurements to update the model, a requirement rarely met in practice. Continuous-time formulations \citep{umlauft2019} avoid the dependence on frequent measurements but require access to the full state and its time derivatives to update the model, which is likewise infeasible in many applications. While state-estimation methods \citep{patwardhan2012} can reconstruct latent states and sometimes estimate selected parameters, they typically either assume known dynamics, lack principled uncertainty quantification, or rely on full-state training data, and therefore do not resolve this limitation. Bayesian time-series models \citep{maiworm2021} bypass the state-space representation, yet they still rely on frequent measurements and make it difficult to incorporate first-principles knowledge, which is naturally expressed in state-space form. Consequently, these approaches are not suitable for settings with infrequent, noisy output measurements, where reconstructing the latent state trajectory is essential for accurate and reliable predictions.

Recent work on uncertainty-aware control has begun to address the dependence of Bayesian state-space models on full-state measurements by placing priors over the dynamics and latent state trajectory, and jointly inferring both from output data \citep{lefringhausen2024,lefringhausen2025}. Since the resulting posterior distribution is analytically intractable, these approaches rely on Markov chain Monte Carlo (MCMC) sampling to update the model and use the resulting samples in the control design. However, existing formulations are restricted to discrete-time systems with measurements available at every time step. They therefore cannot be applied when measurements are infrequent, leaving a clear gap in uncertainty-aware optimal control when only infrequent, noisy output measurements are available.

This work addresses this gap by proposing a framework for uncertainty-aware optimal control of nonlinear systems from infrequent and noisy output measurements. We formulate a prior over the continuous-time dynamics and latent state trajectory in state-space form, enabling the incorporation of domain knowledge. To infer the corresponding posterior distribution, we employ a targeted Metropolis--Hastings sampler that uses numerical ODE integration for likelihood evaluation. While similar ODE-based MCMC schemes have been proposed for parameter identification \citep{girolami2008}, this work exploits the resulting posterior samples for control by constructing a scenario-based optimal control problem in which model uncertainty is explicitly represented. The discretized problem is solved using standard numerical optimization tools. The effectiveness of the proposed approach is demonstrated in a numerical case study on glucose regulation for a Type 1 diabetes patient.

The remainder of the paper is structured as follows. Section~\ref{sec:problem_formulation} formalizes the problem setting. Section~\ref{sec:MH} introduces the Metropolis--Hastings algorithm used for Bayesian inference. Section~\ref{sec:scenario_ocp} presents the scenario-based optimal control formulation. Section~\ref{sec:simulation} evaluates the approach in a numerical example involving a glucose--insulin dynamics model, and Section~\ref{sec:conclusion} concludes the paper.


\section{Problem Formulation}
\label{sec:problem_formulation}
Consider the nonlinear continuous-time system
\begin{subequations} \label{eq:system}
	\begin{align} 
		\dot{\bm{x}}(t) &= \bm{f}_{\bm{\theta}}(\bm{x}(t), \bm{u}(t), t)\\
		\bm{y}(t) &= \bm{g}_{\bm{\theta}}(\bm{x}(t), \bm{u}(t), t) + \bm{w}(t),
	\end{align}
\end{subequations}
with state~$\bm{x}(t) \in \mathbb{R}^{n_x}$, which is not directly observed, input~$\bm{u}(t) \in \mathbb{R}^{n_u}$, and output~$\bm{y}(t) \in \mathbb{R}^{n_y}$. The measurements are affected by noise~$\bm{w}(t)$, where for each fixed sampling time $t_m$ the noise realization $\bm{w}(t_m)$ is drawn from a distribution~$\mathcal{W}_{\bm{\theta}}(t_m)$ with density~$p_{\mathcal{W}}(\bm{w}(t_m)\mid\bm{\theta}, t_m)$. The true dynamics function~$\bm{f}_{\bm{\theta}}(\cdot)$, observation function~$\bm{g}_{\bm{\theta}}(\cdot)$, and the noise distribution~$\mathcal{W}_{\bm{\theta}}$ are unknown. We assume, however, that the model structure is known, and that all unknown quantities are collected in the parameter vector~$\bm{\theta} \in \mathbb{R}^{n_\theta}$. Such parameterizations naturally arise in systems governed by physical laws, which often dictate the structure of the dynamics or measurement process, thereby enabling informative and interpretable models.

At the current time~$t = 0$, we assume access to historical measurements of the system. Negative times~$t \in [-T,0]$ refer to this past data window, whereas positive times~\mbox{$t > 0$} correspond to the control horizon. Within the training window~$[-T,0]$, the input trajectory~$\bm{u}(t)$ is known, and $M$~output measurements $\bm{y}_m = \bm{y}(t_m)$ are available at a (possibly irregular) set of sampling instants~$t_{1:M} \subseteq [-T,0]$. This yields the dataset
\begin{equation}\label{eq:data}
	\mathbb{D} = \left\{ \bm{u}(t) \ \text{for}\ t \in [-T,0], \bm{y}_{1:M} \right\},
\end{equation}
which contains the complete input trajectory and the corresponding noisy output measurements.

To enable Bayesian inference of both the dynamics and the latent state trajectory, we make the following standard assumption.
\begin{assumption}
	\label{as:prior}
	Priors~$p(\bm{\theta})$ and $p(\bm{x}(-T))$ are available for the model parameters and for the initial state at the beginning of the training window. The unknown parameters and initial state are drawn from these priors.
\end{assumption}
This assumption is natural in many practical settings: domain knowledge, operational limits, or previous experiments often provide informative priors for the model parameters and for the range of feasible initial states.

The objective is to find an input trajectory~$\bm{u}(t)$ over a finite prediction horizon~$[0,H]$ that minimizes the continuous-time cost functional
\begin{equation}\label{eq:cost}
	J(\bm{u}(\cdot), \bm{x}(\cdot)) = c_f\left(\bm{x}(H)\right) + \textstyle{\int}_{0}^{H} c \left(\bm{u}(t), \bm{x}(t), t\right) dt,
\end{equation}
where $c$ denotes the running cost and $c_f$ denotes the terminal cost. The input trajectory~$\bm{u}(\cdot)$ must be chosen so that the resulting trajectory satisfies the state and input path constraints
\begin{equation}\label{eq:constraints}
	\bm{h}\left(\bm{u}(t), \bm{x}(t), t\right) \leq \bm{0} \quad \forall t \in [0,H].
\end{equation}
This problem represents the ideal formulation, which would be solvable only under perfect knowledge of the system and its initial state. However, because the system evolution must be inferred from limited data, constraint satisfaction cannot be guaranteed deterministically. The objective is therefore to compute input trajectories that explicitly account for model uncertainty, yielding solutions that are expected to remain feasible and to achieve low cost under likely realizations of the system's evolution.

To address this problem, we first infer a posterior distribution over the unknown dynamics and latent state trajectory using a targeted Metropolis--Hastings scheme equipped with a numerical integrator. Posterior samples obtained in this way represent plausible system evolutions consistent with the available data and prior knowledge. These samples are then used to construct a discretized scenario-based approximation of the optimal control problem (\ref{eq:cost}--\ref{eq:constraints}), from which a robust input trajectory is computed.


\section{Metropolis--Hastings Sampling}
\label{sec:MH}
A central step in the proposed framework is the inference of the posterior distribution~$p(\bm{\theta}, \bm{x}(0) \mid \mathbb{D})$, which combines prior information about the unknown quantities with evidence from the infrequently sampled and noisy measurements, thereby reducing the potentially large model uncertainty that would arise if one relied on the prior alone. We specifically target the state~$\bm{x}(0)$ because, due to the Markov property inherent in the state-space representation~\eqref{eq:system}, $\bm{x}(0)$ summarizes all information that is relevant for predicting the system's future evolution. By Bayes' theorem, this posterior distribution is given by the product of the prior distribution and the likelihood function, normalized by the marginal likelihood~$p(\mathbb{D})$. This expression is not available in closed form: evaluating the likelihood requires solving the nonlinear ODE~\eqref{eq:system}, which in general does not admit an analytical solution, while the marginal likelihood additionally requires integrating over the unknown parameters and initial conditions. To address this difficulty, we employ a targeted Metropolis--Hastings (MH) scheme equipped with a numerical ODE integrator to draw samples from the joint posterior over parameters and latent states. This section introduces the sampling-based inference procedure used to obtain these posterior samples.

The key idea of MH sampling \citep{robert2004} is to generate proposals for the uncertain quantities and accept or reject them based on how well they explain the available measurements. Since the latent state trajectory of a continuous-time system is uniquely determined by the parameters and the initial state, the sampler proposes only a pair~$\bm{z} = (\bm{\theta}, \bm{x}(-T))$ rather than a full trajectory. Proposals are drawn from a distribution~$q(\bm{z}' \mid \bm{z}^{[k]})$ whose support covers the priors $p(\bm{\theta})$ and $p(\bm{x}(-T))$, and are accepted or rejected using the Metropolis--Hastings criterion. The complete MH procedure is summarized in Algorithm~\ref{alg:MH_sampler}.

Evaluating the acceptance probability requires computing the likelihood of the measurement sequence under a proposed pair~$\bm{z}'$, i.e., $p(\bm{y}_{1:M} \mid \bm{z}')$. This, in turn, requires propagating the latent state trajectory to the measurement instants $t_{1:M}$. Let
\begin{equation}\label{eq:flow_map}
	\bm{x}(t) = \Phi(t; \bm{\theta}, \bm{x}(t_0), \bm{u}(\cdot))
\end{equation}
denote the flow map of the state equation in~\eqref{eq:system}, that is, the state~$\bm{x}(t)$ reached by integrating the dynamics from the initial state~$\bm{x}(t_0)$ at time $t_0$ to time $t$ under parameters $\bm{\theta}$ and input $\bm{u}(\cdot)$. The predicted state at time~$t_m$ under a proposal~$\bm{z}'$ is then $\bm{x}'(t_m) = \Phi(t_m; \bm{z}', \bm{u}(\cdot))$. Since $\Phi$ is not available in closed form, it is approximated numerically. Let
\begin{equation}
\widehat{\bm{x}}'(t_m) \approx \Phi(t_m; \bm{z}', \bm{u}(\cdot))
\end{equation}
denote the numerical approximation returned by the ODE solver. For each measurement instant, we define the residual
\begin{equation}\label{eq:residual}
    \bm{r}_m(\bm{z}') := \bm{y}_m -\bm{g}_{\bm{\theta}'}\left(\widehat{\bm{x}}'(t_m), \bm{u}(t_m), t_m\right).
\end{equation}
The likelihood of the observed output measurements is then computed as
\begin{equation}\label{eq:likelihood}
    p(\bm{y}_{1:M} \mid \bm{z}', \bm{u}(\cdot)) \approx \prod_{m=1}^{M} p_{\mathcal{W}}\left(\bm{r}_m(\bm{z}') \mid \bm{\theta}', t_m \right).
\end{equation}
The acceptance probability for a proposed pair~$\bm{z}'$ is given by the standard Metropolis--Hastings ratio
\begin{equation}\label{eq:acceptance_probability}
    \alpha = \min \left(1,\frac{p(\bm{y}_{1:M} \mid \bm{z}')p(\bm{z}')q(\bm{z}^{[k]} \mid \bm{z}')}{p(\bm{y}_{1:M} \mid \bm{z}^{[k]})p(\bm{z}^{[k]})q(\bm{z}' \mid \bm{z}^{[k]})}
    \right).
\end{equation}
This ratio compares the posterior plausibility of the proposed and current pair, weighted by the proposal densities. Proposals that better explain the observed data are more likely to be accepted. If a proposal is accepted, it becomes the next state of the Markov chain; otherwise, the chain remains at its current state. Finally, the sampled posterior~$p(\bm{\theta}, \bm{x}(-T) \mid \mathbb{D})$ is mapped to the target posterior~$p(\bm{\theta}, \bm{x}(0) \mid \mathbb{D})$ by propagating the initial state forward to time~$t=0$ using the ODE flow map, which is again approximated using the same numerical integrator employed during likelihood evaluation.

\begin{algorithm}[t]
    \caption{Metropolis--Hastings sampler}
    \label{alg:MH_sampler}
    \begin{algorithmic}[1]
        \Require Observations $\mathbb{D}$, model $\bm{f}_{\bm{\theta}}(\cdot)$, $\bm{g}_{\bm{\theta}}(\cdot)$, $p_{\mathcal{W}}(\cdot \mid \bm{\theta})$, priors $p(\bm{\theta})$, $p(\bm{x}(-T))$, proposal density $q(\bm{z}' \mid \bm{z})$, initial value $\bm{z}^{[1]}$, number of samples $K$, burn-in period $K_b$, thinning parameter $k_d$
        \Ensure $K$ samples from $p(\bm{\theta}, \bm{x}(0) \mid \mathbb{D})$
        \State Initialize: $k \gets 1$
        \State Compute likelihood $p(\bm{y}_{1:M} \mid \bm{z}^{[1]})$ using \eqref{eq:likelihood}
        \While{$k \leq K_b + 1 + (K-1) (k_d+1)$}
            \State Propose $\bm{z}' \sim q(\bm{z}' \mid \bm{z}^{[k]})$
            \State Simulate latent trajectory under $\bm{z}'$ and evaluate $p(\bm{y}_{1:M} \mid \bm{z}')$ using \eqref{eq:likelihood}
            \State Compute acceptance probability $\alpha$ using \eqref{eq:acceptance_probability}
            \If{$\text{Uniform}(0, 1) < \alpha$}
                \State Accept proposal: $k \gets k + 1$, $\bm{z}^{[k]} \gets \bm{z}'$
            \EndIf
        \EndWhile
        \State Discard burn-in and perform thinning
        \State Compute $\bm{x}^{[k]}(0) = \Phi(0; \bm{z}^{[k]}, \bm{u}(\cdot))$ for each sample
    \end{algorithmic}
\end{algorithm}

Under mild regularity assumptions on the dynamics and the proposal distribution, the MH kernel defined using the exact flow map $\Phi$ admits the desired posterior $p(\bm{z}\mid \mathbb{D})$ as its invariant distribution \citep{tierney1994}. In other words, if the latent trajectory were propagated using the exact solution of the ODE, the resulting Markov chain would converge asymptotically to the true posterior, ensuring that the obtained samples are correctly distributed. In practice, however, the flow map $\Phi$ is not available in closed form and must be approximated numerically. When the likelihood is evaluated using numerically propagated states~$\widehat{\bm{x}}(t_m)$ instead of the exact trajectory~$\Phi(t_m;\bm{z})$, the resulting Metropolis--Hastings transition kernel becomes an approximation of the ideal one. This introduces a small bias in the stationary distribution. Nevertheless, the integration error is typically negligible compared with the intrinsic stochastic variability of the sampler, so the approximate chain remains a high-quality surrogate of the exact posterior.

\begin{remark}\label{rem:adaptive_proposal}
    Efficient MH sampling relies critically on a proposal distribution that is well aligned with the geometry of the posterior. In the present setting, the parameters and the initial state jointly determine the latent trajectory used to explain the measurements, which induces strong correlations in their posterior. A proposal that does not reflect this structure typically leads to poor mixing, low acceptance rates, and highly autocorrelated samples \citep{lefringhausen2025}. A practical approach to constructing an effective proposal is staged inference, where the sampler initially uses only a subset of the measurements and gradually increases the number of data points used in the likelihood evaluation. At each stage, the proposal covariance can be adapted based on the empirical covariance of the chain, allowing the proposal to track the evolving posterior geometry and account for the induced parameter--state correlations.
\end{remark}

After running the MH sampler, the raw sequence of generated samples does not yet constitute a representative set from the target posterior. The first samples of the chain depend heavily on the (generally arbitrary) initialization and may not represent the target distribution. To mitigate this effect, the initial~$K_b$ samples are discarded, which is commonly referred to as the burn-in period. In addition, consecutive samples of the MH sampler are typically correlated, since each new proposal is generated conditionally on the previous accepted sample. For the intended use, strongly autocorrelated samples are undesirable, as they may underrepresent the variability of the posterior distribution. To reduce this autocorrelation, we apply thinning: out of every block of~$k_d+1$ samples, only one sample is retained, while the remaining $k_d$ samples are discarded. The thinning parameter $k_d$ must be chosen to balance statistical independence and computational efficiency\,---\,large enough to ensure that the retained samples exhibit acceptably low autocorrelation, but not so large that an excessive number of samples is discarded. The resulting thinned sequence constitutes a set of samples that are approximately independent draws from the target posterior~$p(\bm{\theta},\bm{x}(0)\mid\mathbb{D})$. These posterior samples represent plausible realizations of the system dynamics consistent with the available data and prior knowledge and form the basis for the scenario-based optimal control formulation described in the next section.


\section{Scenario-Based Optimal Control}
\label{sec:scenario_ocp}
In this section, we describe how the posterior samples obtained from the MH sampler in Section~\ref{sec:MH} can be used to compute an input trajectory that is robust to likely realizations of the actual system dynamics. Section~\ref{subsec:continuous_time_OCP} introduces the underlying idea and formulates the scenario-based optimal control problem in continuous time, while Section~\ref{subsec:discrete_time_OCP} presents its numerical discretization and solution using standard nonlinear programming techniques.

\subsection{Scenario-Based Optimal Control in Continuous Time}
\label{subsec:continuous_time_OCP}
The central idea of the proposed approach is to design a control input that remains feasible and performs well across the range of system behaviors that are plausible under the learned posterior distribution. Rather than relying on a single nominal model or adopting overly conservative worst-case assumptions, we directly exploit the posterior samples~$\{ \bm{\theta}, \bm{x}(0)\}^{[1:K]}$ generated by the MH sampler. Each such sample specifies one plausible realization of the model parameters and current state, consistent with the available measurements and prior information. Because the state evolution of a continuous-time system is uniquely determined by its parameters, initial condition, and input trajectory, each posterior sample induces a distinct possible future trajectory. We therefore interpret these samples as scenarios.

In the proposed scenario-based optimal control formulation, all scenarios are propagated forward under a shared control input trajectory. State and input constraints are enforced for every scenario, while the overall objective aggregates the individual scenario costs. This construction yields a control law that is explicitly robustified against the posterior uncertainty.

Using the flow map~\eqref{eq:flow_map}, each posterior sample~$(\bm{\theta}^{[k]}, \bm{x}^{[k]}(0))$ induces, under a candidate control trajectory $\bm{u}(\cdot)$, the scenario trajectory
\begin{equation}
    \bm{x}^{[k]}(t) = \Phi \left(t; \bm{\theta}^{[k]}, \bm{x}^{[k]}(0), \bm{u}(\cdot)\right),
    \qquad t \in [0,H].
\end{equation}
Replacing the unknown true trajectory in (\ref{eq:cost}--\ref{eq:constraints}) with the scenario trajectories gives a scenario-based version of the optimal control problem. In this formulation, the cost is approximated by the empirical average over the $K$ posterior samples, and all constraints are enforced for every scenario. This yields the following scenario-averaged cost functional
{\small
\begin{equation}\label{eq:scenario_cost_continuous}
    J_{\mathrm{sc}}(\bm{u}(\cdot)) := \frac{1}{K} \sum_{k=1}^{K} \left[ c_f\left(\bm{x}^{[k]}(H)\right) + \textstyle{\int}_0^H c\left(\bm{u}(t),\bm{x}^{[k]}(t), t\right)dt \right],
\end{equation}
}
and the continuous-time scenario optimal control problem becomes
\begin{subequations}\label{eq:scenario_OCP_continuous}
\begin{align}
    & \min_{\bm{u}(\cdot)} \quad J_{\mathrm{sc}}(\bm{u}(\cdot))\\
    \text{s.t.} \quad & \forall t \in [0,H], \quad \forall k  \in \mathbb{N}_{\leq K},\nonumber \\
    &\bm{x}^{[k]}(t) = \Phi\left(t; \bm{\theta}^{[k]}, \bm{x}^{[k]}(0), \bm{u}(\cdot)\right), \label{eq:scenario_OCP_dynamic_constraints}\\
    & \bm{h}\left(\bm{u}(t), \bm{x}^{[k]}(t), t\right) \leq \bm{0}. \label{eq:scenario_OCP_scenario_constraints}
\end{align}
\end{subequations}
By construction, this formulation discourages control inputs that lead to constraint violations or poor performance across the posterior samples, thereby yielding input trajectories that are robust with respect to the uncertainty captured by these scenarios.

\subsection{Time Discretization and Numerical Implementation}
\label{subsec:discrete_time_OCP}

Since the scenario-based optimal control problem \eqref{eq:scenario_OCP_continuous} is posed in continuous time, it cannot be solved directly with standard numerical optimization methods. We therefore introduce a time discretization of the dynamics, costs, and constraints, which yields a finite-dimensional nonlinear program suitable for numerical solution. Following the direct multiple-shooting framework \citep{betts2010}, we introduce a control grid
\begin{equation}
    0 = \tau_0 < \tau_1 < \dots < \tau_{N} = H,
\end{equation}
and restrict the control input to be piecewise constant over each interval,
\begin{equation}
    \bm{u}(t) = \bm{u}_n, \qquad t \in [\tau_n, \tau_{n+1}), \quad \forall n \in \mathbb{N}^0_{\leq N},
\end{equation}
so that the collection $\bm{u}_{0:N-1}$ constitutes the decision variables of the discretized problem. 

For each scenario~$k = 1,\dots,K$, the continuous-time dynamics are discretized by numerically propagating the ODE~\eqref{eq:system} over each interval~$[\tau_n,\tau_{n+1})$ using the sampled parameters~$\bm{\theta}^{[k]}$ and initial state~$\bm{x}^{[k]}(0)$. Let
\begin{equation}
    \bm{x}^{[k]}_{n+1} = \Psi\left(\tau_{n+1},\tau_n; \bm{\theta}^{[k]}, \bm{x}^{[k]}_{n}, \bm{u}_n\right), \quad \forall n \in \mathbb{N}^0_{\leq N},
\end{equation}
denote the numerical flow map produced by the ODE integrator, with $\bm{x}^{[k]}_0 = \bm{x}^{[k]}(0)$. The map $\Psi$ is the discrete counterpart of the continuous-time flow $\Phi$ and is computed using the same numerical solver used during the MH inference step.

The continuous-time cost functional is approximated by
\begin{equation}
    \hat J_{\mathrm{sc}} = \frac{1}{K} \sum_{k=1}^{K} \left[ c_f\left(\bm{x}^{[k]}_{N}\right) + \sum_{n=0}^{N-1} c\left(\bm{u}_n, \bm{x}^{[k]}_n, \tau_n\right) \Delta_n \right],
\end{equation}
where $\Delta_n = \tau_{n+1} - \tau_n$ denotes the grid spacing.

The constraints \eqref{eq:constraints} are imposed at all grid points for every scenario. The discretized scenario-based optimal control problem then becomes the nonlinear program
\begin{subequations}\label{eq:scenario_OCP_NLP}
\begin{align}
    \min_{\bm{u}_{0:N-1}} \quad & \hat J_{\mathrm{sc}}, \\
    \text{s.t.} \quad & \forall n \in \mathbb{N}^0_{\leq N}, \quad \forall k  \in \mathbb{N}_{\leq K},\nonumber\\
    & \bm{x}^{[k]}_{n+1} = \Psi\left(\tau_{n+1},\tau_n; \bm{\theta}^{[k]}, \bm{x}^{[k]}_{n}, \bm{u}_n \right), \\
    & \bm{h}\left(\bm{u}_n, \bm{x}^{[k]}_n, \tau_n\right) \leq \bm{0}.
\end{align}
\end{subequations}
The resulting finite-dimensional optimization problem can be solved using standard nonlinear programming solvers; see \citep{nocedal2006}.

The conservatism of the resulting control law is closely tied to the uncertainty represented by the posterior distribution from which the scenarios are drawn. Incorporating additional measurements can reduce this uncertainty and thereby typically leads to less conservative control inputs. Importantly, the proposed framework does not require a minimum number of data points to be applicable: in the absence of measurements, the scenario problem is constructed from prior samples and reflects prior uncertainty alone.

For a given posterior distribution, the robustness of the resulting control law is further affected by several numerical and sampling-related factors. First, since the constraints in \eqref{eq:scenario_OCP_NLP} are imposed only at the discretization nodes $\tau_{0:N}$, they ensure feasibility only at these time points. Even with a perfect model, constraint satisfaction between nodes cannot be guaranteed merely by pointwise enforcement; refining the grid reduces the risk of inter-sample violations but increases the computational burden of the NLP \citep{betts2010}. Second, the extent to which the scenario program captures the posterior uncertainty is governed by both the number and the diversity of the posterior samples. Increasing the number of approximately independent samples generally improves robustness, as it provides a broader and more representative coverage of the posterior distribution, thereby exposing the optimizer to a wider range of plausible system behaviors. However, this comes at the cost of a larger NLP, as the number of scenario-dependent variables and constraints grows linearly with $K$. Conversely, strong autocorrelation in the Markov chain leads to clusters of nearly identical scenarios, artificially narrowing the represented uncertainty and weakening the practical robustness of the computed input trajectory. Finally, the predicted state trajectories rely on numerical integration of the dynamics. For consistency between learning and control, it is advisable to employ the same ODE solver used during posterior inference. Although numerical integration introduces an additional approximation error, this effect is typically less significant than the uncertainty encoded in the posterior samples and the discretization error associated with the finite control grid.

It is worth noting that, under the idealized assumption of perfect numerical integration, one could formalize this robustness and provide probabilistic guarantees of constraint satisfaction at the grid points. Such guarantees may be obtained by validating the computed input trajectory on an additional set of posterior samples or, alternatively, by adopting a robust cost formulation and invoking results from robust optimization; closely related arguments have been established in \citep{lefringhausen2024}. However, this lies beyond the scope of the present work. Instead, the practical robustness of the proposed approach is demonstrated in the numerical case study of the following section.

\section{Simulation}
\label{sec:simulation}
We evaluate the proposed framework on blood glucose regulation in patients with Type~1 diabetes. This setting provides a natural test case: the dynamics are nonlinear and patient-specific, the physiological states are only partially observed through glucose measurements, observations may occur at irregular times, and the control task is safety-critical.

\subsection{Setup}
We employ an extended Bergman minimal model~\citep{bergman1981,ali2011} with three states\,---\,blood glucose~$G(t)$~[\si{mg/dL}], plasma insulin~$I(t)$~[\si{mU/L}], and remote insulin action~$X(t)$~[\si{1/min}]\,---\,driven by exogenous insulin input~$u(t)$ and known meal disturbances. Only glucose is directly measured and is corrupted by Gaussian measurement noise with standard deviation~$\sigma = \SI{8}{mg/dL}$; the remaining states are latent. The system is simulated over 18~hours (6\,am--12\,am) with three meals. The first 12~hours serve as training data, and the remaining 6~hours constitute the control horizon. We estimate the physiological parameters~$p_2$, $p_3$, and $n$\,---\,the remaining parameters are assumed known\,---\,together with the latent state trajectory from $M=200$ glucose measurements sampled at random times within the training window.

For inference, independent lognormal priors are placed on the unknown physiological parameters, and Gaussian priors are placed on the initial state. Both the MH sampler and the OCP use a fourth-order Runge--Kutta integrator (RK4) with a step size of \SI{0.5}{min}. To determine an appropriate thinning interval, we run the MH sampler for $10^5$ iterations on a representative instance and select a thinning factor of $k_d = 25$ such that the retained samples exhibit negligible empirical autocorrelation.

The control objective is to maintain glucose near the target of \SI{80}{mg/dL}. The controller is subject to the glucose safety bounds \mbox{$70 \leq G(t) \leq \SI{180}{mg/dL}$} and to the insulin pump limits \mbox{$0 \leq u(t) \leq \SI{20}{mU/min}$}. The cost functional penalizes deviations from the glucose target and control effort. The OCP is solved using JuMP~\citep{lubin2023} with IPOPT~\citep{wachter2006} and HSL linear solvers~\citep{hsl2025}.

To assess the robustness of the proposed method with respect to variability in patient dynamics and initial conditions, we conduct a Monte Carlo study consisting of 100 independent simulation runs. In each run, the true physiological parameters, as well as the initial state at 6\,am, are sampled from the prior distributions. For each sampled instance, measurements are generated over the training window, and the MH sampler is used to obtain $K = 100$ posterior samples. These samples are then used to formulate the optimal control problem described in Section~\ref{sec:scenario_ocp}. The resulting control input is then applied to the corresponding true system realization to evaluate the system response.

For comparison, we also evaluate a nominal-model baseline. In this approach, the parameters are fixed at the geometric mean of their respective prior distributions, and the latent state is estimated using an extended Kalman filter (EKF) equipped with the same RK4 integration scheme to accommodate infrequent measurements. The optimal input trajectory is then computed using this nominal model and the EKF state estimate as the initial condition.

The implementation used for the numerical experiments is available at \href{https://github.com/TUM-ITR/ode-mh-planner}{https://github.com/TUM-ITR/ode-mh-planner}. Further details on the experimental setup, including the prior specifications and additional results are provided in the accompanying \href{https://tum-itr.github.io/ode-mh-planner/dev/experiments/overview/}{documentation}.

\newpage
\subsection{Results}
Table~\ref{tab:results} shows that the nominal-model EKF baseline violates the prescribed glucose safety bounds of $70$--$\SI{180}{mg/dL}$ in 44 of the 100 runs, highlighting its inability to maintain physiologically safe glucose levels under uncertainty. In contrast, the proposed MH-scenario approach satisfies all constraints in every run, demonstrating strong robustness to uncertainty in both the model parameters and the latent state. The proposed method also achieves substantially lower costs. By jointly inferring the state and the model parameters and optimizing over multiple posterior scenarios, it obtains more accurate and reliable predictions than a single nominal model, resulting in more effective control actions and improved overall performance.

\begin{table}[t]
\centering
\caption{\small Cost and constraint violations over 100 Monte Carlo runs.}
\label{tab:results}
\vspace{-0.1cm}
\begin{tabular}{lcc}
\toprule
Method & Cost (×10\textsuperscript{4}) & Violations \\
\midrule
MH-scenario & $8.98 \pm 3.50$  & 0/100 \\
Nominal + EKF  & $16.55 \pm 13.27$  & 44/100 \\
\bottomrule
\end{tabular}
\end{table}

\begin{figure}[t]
		\pgfplotsset{width=9cm, compat = 1.18, 
			height = 5cm, grid= major, 
			legend cell align = left, ticklabel style = {font=\tiny},
			every axis label/.append style={font=\tiny}
		}
		\def\file{data/optimal_control.txt}
		
		\centering
		\begin{tikzpicture}
			\begin{axis}[
				grid=both,
				xmin=0, xmax=360,
				ymin=55, ymax=140,
				xtick distance=60,
				ytick distance=20,
				ylabel={$G$~[\si{mg/dL}]}, 
				xlabel={$t$~[\si{min}]},
				set layers=standard,
				reverse legend,
				legend style={font=\tiny, at={(1,1)},anchor=north east, row sep=0pt},
				ylabel shift = -4 pt,
				xlabel shift = -4 pt]
				
				\draw[fill=red, fill opacity=0.2,red, opacity=0.2] (0,50) rectangle (360,70);
				\draw[fill=red, fill opacity=0.2,red, opacity=0.2] (0,180) rectangle (360,190); 
				\addlegendimage{area legend, fill=red, fill opacity=0.2, draw=none}
				\addlegendentry{Safety bounds}
				
				\addplot[ultra thick, Orange] table[x=t,y=x1_nominal]{\file};
				\addlegendentry{Nominal (realized)}
				
				\addplot[ultra thick, dashed, blue] table[x=t,y=x1_MH]{\file};
				\addlegendentry{MH (realized)}
				
				\addplot[name path=A, forget plot, thick, opacity=0.2] table[x=t,y=x1_min]{\file};
				\addplot[name path=B, thick, opacity=0.2] table[x=t,y=x1_max]{\file};
				\tikzfillbetween[of=A and B]{opacity=0.2};
				\addlegendentry{MH (scenarios)}
				
				\addplot[ultra thick,OliveGreen] table[x=t,y=x1_mean]{\file};
				\addlegendentry{MH (mean)}
			\end{axis}
		\end{tikzpicture}
        \vspace*{-0.3cm}
        
		\caption{\small Glucose trajectories over the control horizon for one representative run. The green curve shows the mean prediction of the MH-scenario planner, and the gray band denotes the envelope of the 100 posterior scenarios used for planning. The dashed blue curve depicts the realized trajectory obtained when applying the MH-scenario control input, while the solid orange curve shows the realized trajectory under the Nominal + EKF baseline. The red bands indicate the prescribed safety bounds.}
		\label{fig:single_run_trajectory}
		\vspace*{0.1cm}
	\end{figure}
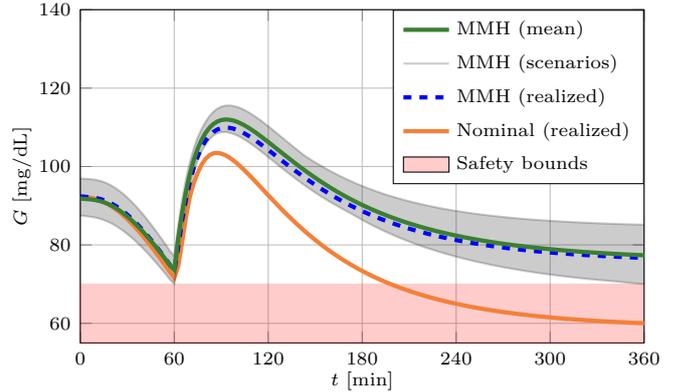

Figure~\ref{fig:single_run_trajectory} shows one representative run. The MH-scenario controller anticipates the meal at 7\,pm by increasing insulin delivery beforehand, and the realized trajectory remains within the safety bounds and inside the scenario envelope. In contrast, the nominal baseline drives glucose to approximately \SI{60}{mg/dL}, violating the lower safety bound.

\section{Conclusion}
\label{sec:conclusion}
This paper presents a method for uncertainty-aware optimal control of nonlinear continuous-time systems from infrequent and noisy output measurements. A prior over the dynamics in state-space representation is updated using a Metropolis--Hastings sampler, and the resulting samples are used to formulate a scenario optimal control problem. Relevant directions for future work include the derivation of formal probabilistic guarantees of constraint satisfaction and the incorporation of process noise in the continuous-time dynamics, which would require numerical approximations of the stochastic state transitions between measurement times to generalize the MH sampler to a particle marginal MH scheme.

\section*{Declaration of Generative AI and AI-Assisted Technologies in the Writing Process}
During the preparation of this work, the authors utilized ChatGPT to enhance language and readability and to aid in programming the simulation. After using this tool, the authors reviewed and edited the content as needed and take full responsibility for the content of the publication.

\bibliography{references}

\end{document}